

\newcount\driver \newcount\mgnf \newcount\tipi
\mgnf=0
\driver=1
\tipi=2
\newskip\ttglue
\def\TIPITOT{
\font\dodicirm=cmr12
\font\dodicii=cmmi12
\font\dodicisy=cmsy10 scaled\magstep1
\font\dodiciex=cmex10 scaled\magstep1
\font\dodiciit=cmti12
\font\dodicitt=cmtt12
\font\dodicibf=cmbx12
\font\dodicisl=cmsl12
\font\ninerm=cmr9
\font\ninesy=cmsy9
\font\eightrm=cmr8
\font\eighti=cmmi8
\font\eightsy=cmsy8
\font\eightbf=cmbx8
\font\eighttt=cmtt8
\font\eightsl=cmsl8
\font\eightit=cmti8
\font\seirm=cmr6
\font\seibf=cmbx6
\font\seii=cmmi6
\font\seisy=cmsy6
\font\dodicitruecmr=cmr10 scaled\magstep1
\font\dodicitruecmsy=cmsy10 scaled\magstep1
\font\tentruecmr=cmr10
\font\tentruecmsy=cmsy10
\font\eighttruecmr=cmr8
\font\eighttruecmsy=cmsy8
\font\seventruecmr=cmr7
\font\seventruecmsy=cmsy7
\font\seitruecmr=cmr6
\font\seitruecmsy=cmsy6
\font\fivetruecmr=cmr5
\font\fivetruecmsy=cmsy5
\textfont\truecmr=\tentruecmr
\scriptfont\truecmr=\seventruecmr
\scriptscriptfont\truecmr=\fivetruecmr
\textfont\truecmsy=\tentruecmsy
\scriptfont\truecmsy=\seventruecmsy
\scriptscriptfont\truecmr=\fivetruecmr
\scriptscriptfont\truecmsy=\fivetruecmsy
\def \ottopunti{\def\rm{\fam0\eightrm}
\textfont0=\eightrm \scriptfont0=\seirm \scriptscriptfont0=\fiverm
\textfont1=\eighti \scriptfont1=\seii   \scriptscriptfont1=\fivei
\textfont2=\eightsy \scriptfont2=\seisy   \scriptscriptfont2=\fivesy
\textfont3=\tenex \scriptfont3=\tenex   \scriptscriptfont3=\tenex
\textfont\itfam=\eightit  \def\it{\fam\itfam\eightit}%
\textfont\slfam=\eightsl  \def\sl{\fam\slfam\eightsl}%
\textfont\ttfam=\eighttt  \def\tt{\fam\ttfam\eighttt}%
\textfont\bffam=\eightbf  \scriptfont\bffam=\seibf
\scriptscriptfont\bffam=\fivebf  \def\bf{\fam\bffam\eightbf}%
\tt \ttglue=.5em plus.25em minus.15em
\setbox\strutbox=\hbox{\vrule height7pt depth2pt width0pt}%
\normalbaselineskip=9pt
\let\sc=\seirm  \let\big=\eightbig  \normalbaselines\rm
\textfont\truecmr=\eighttruecmr
\scriptfont\truecmr=\seitruecmr
\scriptscriptfont\truecmr=\fivetruecmr
\textfont\truecmsy=\eighttruecmsy
\scriptfont\truecmsy=\seitruecmsy
}\let\nota=\ottopunti}

\newfam\msbfam   
\newfam\truecmr  
\newfam\truecmsy 

\def\TIPI{
\font\eightrm=cmr8
\font\eighti=cmmi8
\font\eightsy=cmsy8
\font\eightbf=cmbx8
\font\eighttt=cmtt8
\font\eightsl=cmsl8
\font\eightit=cmti8
\font\tentruecmr=cmr10
\font\tentruecmsy=cmsy10
\font\eighttruecmr=cmr8
\font\eighttruecmsy=cmsy8
\font\seitruecmr=cmr6
\textfont\truecmr=\tentruecmr
\textfont\truecmsy=\tentruecmsy
\def \ottopunti{\def\rm{\fam0\eightrm}
\textfont0=\eightrm
\textfont1=\eighti
\textfont2=\eightsy
\textfont3=\tenex \scriptfont3=\tenex   \scriptscriptfont3=\tenex
\textfont\itfam=\eightit  \def\it{\fam\itfam\eightit}%
\textfont\slfam=\eightsl  \def\sl{\fam\slfam\eightsl}%
\textfont\ttfam=\eighttt  \def\tt{\fam\ttfam\eighttt}%
\textfont\bffam=\eightbf
\def\bf{\fam\bffam\eightbf}%
\tt \ttglue=.5em plus.25em minus.15em
\setbox\strutbox=\hbox{\vrule height7pt depth2pt width0pt}%
\normalbaselineskip=9pt
\let\sc=\seirm  \let\big=\eightbig  \normalbaselines\rm
\textfont\truecmr=\eighttruecmr
\scriptfont\truecmr=\seitruecmr
}\let\nota=\ottopunti}
\def\TIPIO{
\font\setterm=amr7 
\font\settesy=amsy7 \font\settebf=ambx7 
\def \settepunti{\def\rm{\fam0\setterm}
\textfont0=\setterm   
\textfont2=\settesy   
\textfont\bffam=\settebf  \def\bf{\fam\bffam\settebf}
\normalbaselineskip=9pt\normalbaselines\rm
}\let\nota=\settepunti}


\newskip\ttglue
\ifnum\tipi=0\TIPIO \else\ifnum\tipi=1 \TIPI\else \TIPITOT\fi\fi

%
%

\newdimen\xshift \newdimen\xwidth \newdimen\yshift
%
%
\def\ins#1#2#3{\vbox to0pt{\kern-#2 \hbox{\kern#1 #3}\vss}\nointerlineskip}
%
%
%
\def\insertplot#1#2#3#4#5{\par%
\xwidth=#1 \xshift=\hsize \advance\xshift by-\xwidth \divide\xshift by 2%
\yshift=#2 \divide\yshift by 2%
\line{\hskip\xshift \vbox to #2{\vfil%
\ifnum\driver=0 #3
\special{ps: plotfile #4.ps} 
\fi
\ifnum\driver=1 #3 \includegraphics{#4.ps}\fi
\ifnum\driver=2 #3
\ifnum\mgnf=0\special{#4.ps 1. 1. scale} \fi
\ifnum\mgnf=1\special{#4.ps 1.2 1.2 scale}\fi
 \fi }\hfill}\line{\hfil\hbox{#5}\hfil}\vskip.3truecm}

\def\initfig#1{%
\catcode`\%=12\catcode`\{=12\catcode`\}=12
\catcode`\<=1\catcode`\>=2
\openout13=#1.ps}
\def\endfig{%
\closeout13
\catcode`\%=14\catcode`\{=1
\catcode`\}=2\catcode`\<=12\catcode`\>=12}


\ifnum\mgnf=0
\magnification=\magstep0\hoffset=0.cm
\voffset=-1truecm\hoffset=-.5truecm\hsize=16.5truecm \vsize=24.truecm
\baselineskip=14pt  
\parindent=12pt
\lineskip=4pt\lineskiplimit=0.1pt      \parskip=0.1pt plus1pt
\def\st{\scriptstyle}\def\sst{\scriptscriptstyle}

\fi
\ifnum\mgnf=1
\magnification=\magstep1\hoffset=0.cm
\voffset=-1truecm\hoffset=-.5truecm\hsize=16.5truecm \vsize=24.truecm
\baselineskip=14pt  
\parindent=12pt
\lineskip=4pt\lineskiplimit=0.1pt      \parskip=0.1pt plus1pt
\def\st{\scriptstyle}\def\sst{\scriptscriptstyle}

\fi

 \let\b=\beta  \let\c=\chi \let\d=\delta  \let\e=\varepsilon
\let\f=\varphi \let\g=\gamma \let\h=\eta    \let\k=\kappa  \let\l=\lambda
\let\m=\mu   \let\n=\nu   \let\o=\omega    \let\p=\pi  
\let\r=\rho  \let\s=\sigma \let\t=\tau   
  \let\z=\zeta  
 \let\F=\Phi  \let\G=\Gamma  \let\L=\Lambda 
     \let\Si=\Sigma


\def\data{\number\day/\ifcase\month\or gennaio \or febbraio \or marzo \or
aprile \or maggio \or giugno \or luglio \or agosto \or settembre
\or ottobre \or novembre \or dicembre \fi/\number\year;\,\the\time}


\setbox200\hbox{$\scriptscriptstyle \data $}

\newcount\pgn \pgn=1
\def\foglio{\number\numsec:\number\pgn
\global\advance\pgn by 1}
\def\foglioa{A\number\numsec:\number\pgn
\global\advance\pgn by 1}

%


\global\newcount\numsec\global\newcount\numfor
\global\newcount\numfig
\gdef\profonditastruttura{\dp\strutbox}

\def\senondefinito#1{\expandafter\ifx\csname#1\endcsname\relax}

\def\SIA #1,#2,#3 {\senondefinito{#1#2}%
\expandafter\xdef\csname #1#2\endcsname{#3}\else
\write16{???? ma #1,#2 e' gia' stato definito !!!!} \fi}

\def\etichetta(#1){(\veroparagrafo.\veraformula)%
\SIA e,#1,(\veroparagrafo.\veraformula) %
\global\advance\numfor by 1%
\write15{\string\FU (#1){\equ(#1)}}%
\write16{ EQ #1 ==> \equ(#1) }}

\def\FU(#1)#2{\SIA fu,#1,#2 }

\def\etichettaa(#1){(A\veraformula)%
\SIA e,#1,(A\veraformula) %
\global\advance\numfor by 1%
\write15{\string\FU (#1){\equ(#1)}}%
\write16{ EQ #1 ==> \equ(#1) }}

\def\getichetta(#1){Fig. \verafigura
\SIA e,#1,{\verafigura} %
\global\advance\numfig by 1%
\write15{\string\FU (#1){\equ(#1)}}%
\write16{ Fig. \equ(#1) ha simbolo  #1  }}

\newdimen\gwidth

\def\BOZZA{
\def\alato(##1){%
 {\vtop to \profonditastruttura{\baselineskip
 \profonditastruttura\vss
 \rlap{\kern-\hsize\kern-1.2truecm{$\scriptstyle##1$}}}}}
\def\galato(##1){\gwidth=\hsize \divide\gwidth by 2%
 {\vtop to \profonditastruttura{\baselineskip
 \profonditastruttura\vss
 \rlap{\kern-\gwidth\kern-1.2truecm{$\scriptstyle##1$}}}}}
\footline={\rlap{\hbox{\copy200}\ $\st[\number\pageno]$}\hss\tenrm
\foglio\hss}
}

\def\alato(#1){}
\def\galato(#1){}

\def\veroparagrafo{\number\numsec}\def\veraformula{\number\numfor}
\def\verafigura{\number\numfig}

\def\geq(#1){\getichetta(#1)\galato(#1)}
\def\Eq(#1){\eqno{\etichetta(#1)\alato(#1)}}
\def\eq(#1){\etichetta(#1)\alato(#1)}
\def\Eqa(#1){\eqno{\etichettaa(#1)\alato(#1)}}
\def\eqa(#1){\etichettaa(#1)\alato(#1)}
\def\eqv(#1){\senondefinito{fu#1}$\clubsuit$#1
\write16{#1 non e' (ancora) definito}%
\else\csname fu#1\endcsname\fi}
\def\equ(#1){\senondefinito{e#1}\eqv(#1)\else\csname e#1\endcsname\fi}

\def\include#1{
\openin13=#1.aux \ifeof13 \relax \else
\input #1.aux \closein13 \fi}
\openin14=\jobname.aux \ifeof14 \relax \else
\input \jobname.aux \closein14 \fi
\openout15=\jobname.aux

%



\let\ciao=\bye 
\def\pagina{{\vfill\eject}} \def\\{\noindent}
\def\bra#1{{\langle#1|}} \def\ket#1{{|#1\rangle}}
\def\media#1{{\langle#1\rangle}}
\def\ie{\hbox{\it i.e.\ }}

  \let\ig=\int \let\io=\infty  

\let\dpr=\partial \def\V#1{\vec#1}   
\def\oo{{\V\o}}   
\def\xx{{\V x}} \def\yy{{\V y}} \def\kk{{\V k}}

\def\tende#1{\vtop{\ialign{##\crcr\rightarrowfill\crcr
              \noalign{\kern-1pt\nointerlineskip}
              \hskip3.pt${\scriptstyle #1}$\hskip3.pt\crcr}}}
\def\otto{{\kern-1.truept\leftarrow\kern-5.truept\to\kern-1.truept}}

\def\mbox{\hbox}\def\lis#1{{\overline#1}}
\def\EE{{\cal E}}

  \def\LL{{\cal L}} \def\DD{{\cal D}}
   
\def\fra#1#2{{#1\over#2}}
\def\ap{{\it a priori\ }}

\def\={{\equiv}}

\let\0=\noindent
\def\*{\vglue0.3truecm}

\def\pf{p_{\sst F}}

\vglue1.cm
\centerline{\bf Disorder in the 1D spinless Holstein model.}
\vglue1.cm
\centerline{G. Benfatto,\footnote{${}^1$}{\nota Dipartimento di
Matematica, $II^a$ Universit\`a di Roma, Via della Ricerca Scientifica,
00133, Roma, Italia.} G. Gallavotti,\footnote{${}^2$}{\nota
Dipartimento di Fisica, $I^a$ Universit\`a di Roma, P.le Moro 2, 00153
Roma, Italia.}J. L. Lebowitz\footnote{${}^3$}{\nota Dept. of Mathematics
and Physics, Rutgers University, New Brunswick, NJ 08903, USA.}}
\vglue1.5cm
\0{\it Abstract: We investigate a spinless fermion system on a one dimensional
lattice interacting locally with the optical modes of a quantized phonon
field: the Holstein model. The system is shown to have a disordered
ground state, for small enough coupling, at any density.  This is in
contrast to the non quantized phonon case, the static Holstein model,
which at half filling has an ordered ground state for all couplings.}
\vglue1.cm
\0{\it\S 1 Model and results.}
\vskip0.5cm\numsec=1\numfor=1
The hamiltonian is the sum of a free lattice fermion hamiltonian $H_F$,
a free phonon field hamiltonian $H_B$ and a local fermion phonon
interaction $V$:
$$\eqalign{
H=&\left\{ -\fra{\hbar^2}{2m l}\sum_\xx [a^+_\xx a^-_{\xx+l} +a^+_\xx
a^-_{\xx-l}+(\m_0-2)n_\xx] \right\}+\cr
&+\left\{ -\fra{\hbar^2}{2\s l}\sum_\xx\fra{\dpr^2}{\dpr\f_\xx^2}\,
+\,\fra{l\s}2\sum_\xx [\o^2\f_\xx^2
+\fra{c^2}{l^2}(\f_\xx-\f_{\xx+l})^2] \right\}+\cr
&+\left\{ \fra{\hbar^2}{2m l}\sum_\xx
(-\n+ \l\f_\xx\,)(n_\xx -\fra12) \right\}\cr}\Eq(1.1)$$
where $\xx$ is a point on the lattice with spacing $l$, $\xx=j l,\ j\in
(-\fra{L}{2l},\fra{L}{2l}]$, $j$ integer. Periodic boundary conditions
are imposed at $\pm \fra{L}2$ (by identifying such points). The
$a^\pm_\xx$ are creation and annihilation operators for a spinless
fermion at $\xx$, $n_\xx=a^+_\xx a^-_\xx$ and $a^+_x a^-_y + a^-_y a^+_x
= \d_{xy}$.

The field $\f_\xx$ represents a quantized bosonic field, corresponding
to a discretized vibrating string with linear density $\s$, optical
frequency $\o$ and maximum wave propagation speed $c$. The physical
meaning of $\f_\xx$ is that of deformation of the crystal cell sitting
at $\xx$.\\ The parameter $m$ is a scale parameter fixing the bare
fermion mass. The fermions chemical potential is $\m=\m_0+\n$.

The reason for writing $\m$ in this form is that, since the early works
on the theory of Fermi systems, [LW], it has been realized that it is
more natural to study the properties of such systems when the
interaction strength $\l$ is varied at fixed Fermi momentum rather than
at fixed $\m$.  For the free particle system, corresponding to $\l=0$,
the Fermi momentum $\pf$ is defined as the value of the momentum where
the momentum distribution, which is the Fourier transform of the one
particle reduced density matrix or equal time pair Schwinger function,
has a discontinuity. This manifests itself in the one particle density
matrix having an oscillatory decay $G_0(\xx)\simeq |\xx|^{-1}\sin \pf
|\xx|$ and $\pf$ is obtained from $\m$ by the relation $\m=2(1-\cos
l\pf)$.  For $\l\ne0$ a Fermi momentum $P(\m,\l)$ can still be defined
via the position of the singularity of the Fourier transform of the one
particle reduced density matrix. In particular, for $\l$ small we shall
prove that the one point reduced density matrix has an oscillating
leading asymptotic behaviour $G(\V x)$ proportional to $|\V
x|^{-1-2\h}(O(\l)+\sin P\V |x|)$ for some $\h>0$.  We now define
$\m_0\=2(1-\cos l \pf)$ and choose $\n$ such that $P(\m,\l)=\pf$.  This
defines $\n$ as a function of $\pf$ and $\l$ with $\n = 0$ when $\l =
0$.

It follows from our analysis (see [BGPS] and below) that $P$ is smooth
in $\m,\l$ for $\m$ near any prefixed $\m_0\in (0,2)$ and $\l$ small
enough (how small depends on the value of $\m_0$).  Therefore we can
write $P=\pf+c_1 (\m-\m_0)+b_1\l+\ldots$ and $c_1\ne0$ (in fact
$c_1=(2l\sin l\pf)^{-1}$). Setting $P=\pf$ then yields
$\n=d_1\l+\ldots$ with $d_1=b_1/c_1$.

We note that, according to the (formal) Luttinger theorem [L], fixing
$P(\m,\l)$ is equivalent to fixing the physical density
$\r=\r(\m_0+\n,\l)$, \ie $P$ is independent of $\l$ if $\r$ is fixed: in
fact $P=\p\r$. The (formal) extension of this theorem (proved formally
in [LW], [L] and formulated there as the "conservation of the Fermi
surface volume" at constant density) to cover the present case is
discussed, for completeness, in the appendix.

In general the value of $\n$ is a complicated function of $\l$, which
can only be determined order by order in perturbation theory. There is
however an exception; in fact, setting $\e_\xx=(-1)^{\xx/l}$, the
unitary transformation $a^\pm_\xx\to \e_\xx a^\mp_\xx$,
$\f_\xx\to-\f_\xx$ maps the hamiltonian with chemical potential $\m=2$
and $\n=\n_0$ into that with $\m=2$ and $\n=-\n_0$ and a state with
density $\r$ into one with density $1-\r$. Hence we see that, if $\n=0$,
there must be a ground state with density $\fra12$. Furthermore the
hamiltonian has other symmetries; namely translation invariance and
reflection (parity) symmetry. Thus, if we suppose that the ground state
is unique (a property that we expect but do not prove) then by applying
to it the above three symmetries we see that $G(\V x)=\media{a^-_{\V x}
a^+_0}=0$ for $\V x/l$ even. Hence if one can prove, as we claim here,
the existence of $P$ such that $G(\V x)$ is proportional to $|\V
x|^{-1-2\h}(O(\l)+\sin P\V |x|)$, it follows that $P=\p/2l$ (otherwise,
for $\xx/l$ even and large enough, $\sin P\xx$ could be of order $1$),
so that the Luttinger theorem is automatically satisfied.

The units can be fixed so that:

$$\hbar=m=l=m\s\o^2=1\Eq(1.3)$$

\0Setting $b=c\o^{-1}, \s_0^2=\s$, $H$ becomes:

$$\eqalign{ 2H=&\left[\sum_\xx (-a^+_\xx a^-_{\xx+1}-a^+_\xx
a^-_{\xx-1} +(2-\m)a^+_\xx a^-_\xx)\right]+\cr
&+\left\{-\fra1{\s_0^2}\sum_\xx\fra{\dpr^2}{\dpr
\f_\xx^2}\,+\,\sum_\xx(\f_\xx^2+
b^2\,(\f_\xx-\f_{\xx+1})^2)\right\}+\cr &+\left(\sum_\xx
(-\n+\l\f_\xx)\,(a_\xx^+ a^-_\xx-\fra12)\right)\cr}\Eq(1.4)$$

\0with the half filled band case corresponding to $\m=2,\n=0$.

\vglue.5cm
\0{\it\S 1.1 The Static Holstein Model.}
\vskip0.25cm

If $\s_0=+\io,\ b=0$ the model \equ(1.4) is the static Holstein model
with ``magnetic field'' equal to $\l/4$, [LM]. The ground state problem
is now equivalent to the computation of the fermionic energy
$E(\l,\n;\f)$ in the presence of the external field $\f_\xx$ and then
finding the field $\f$ minimizing:

$$E(\l,\n;\f)+\fra12\sum_\xx \f_\xx^2\Eq(1.5)$$

\0Calling $\EE(\l,\n)$ the minimum value, the corresponding density is
obtained by setting $\r-\fra12=-2 \dpr_\n\EE(\l,\n)$ which, if $\r$ is
given \ap, is an equation determining $\n$ as a function of $\l,\r$.

The case $\m=2,\n=0,\r=\fra12$, the so called ``half filled band'' case,
has been solved by [LM], following the methods used by [KL] for the case
in which the field $\f_\xx$ can only take values $\pm1$ (the
Falikov--Kimball model, see also [MM]). It is shown in [LM] that the
minimizing field is $\f_\xx=(-1)^\xx f$ where $f$ is a suitable
constant, which can be easily computed by remarking that
$E(\l,\n;(-1)^\xx f)$ can be evaluated by the Bloch waves techniques.

One finds that $\l\ne0$ implies $f\ne0$. This means that, if
$\l\ne0$, there are two non translation invariant ground states, in
which the field $\f_{\V x}$ has a periodic structure with period $2$.
This is interpreted by saying that at half filling the
atoms of the system acquire a crystalline ordering with a period $2$, at
any non zero coupling strength; \ie the Peierls instability.
They, thus, behave as if they were non interacting particles
immersed in a periodic potential with period $2$.

It follows from this that the one particle energies are split into two
bands, the first corresponding to the momenta $|k|<\fra\p2$, and the
second to the larger $k's$. The two bands are separated by a gap
$\k(\l)$ and, as a consequence, the fermions are in a filled band state
(because the Fermi momentum is $\pf=\fra\p2$ for $\r=1/2$, by the above
symmetry). Consequently the two point equal time Schwin\-ger function
for the fermions, $G(\xx)=\lim_{t\to 0^-} G(\xx,t)$, decays at large
separation as,
$$|G(\xx)|\simeq e^{-\k(\l)|\xx|}\Eq(1.6)$$
where $\k(\l)$ is the energy gap at momentum $\pf=\p/2$. Note that
the energy gap $\k(\l)\tende{\l\to0}0$, and that at $\l=0$ one has instead:
$G(\xx) = G_0(\xx) = -(\p \xx)^{-1}\sin \pf\xx$.

\vglue.5cm
\0{\it\S 1.2 The Dynamic Holstein Model.}
\vskip0.25cm

The above picture holds {\it for all} values of $\l\ne0$ at half filling
($\r=\fra12,\n=0$) when the phonon field is treated classically, by
setting $\s_0^{-2}=0$ in (1.3) from the beginning.  We shall show that
the results in [BGPS] imply that, if $\s_0<\io$, \ie if quantum effects
are not neglected, then for {\it any density} $\r\in (0,1)$, and $\l$
small enough (depending on $\s_0,b$), the decay of $G(\xx,0)$ is given
by:
$$G(\xx,0)\simeq \fra{G_0(\xx,0)}{|\xx|^{2\h(\l)}}\Eq(1.7)$$
with $\h(\l)$ analytic in $\l^2$; $\h(\l)=a \l^4+O(\l^6)$ if $b\ne0$,
or $\h(\l)=a'\l^6+O(\l^8)$ if $b=0$, with $a,a'\ne0$.

This is clearly incompatible with a periodic minimizing state of the
field $\f_x$, showing that the Holstein model ground state for a
spinless fermion system is ``disordered'' at small coupling, \ie there
is no long range order in $<\f_\xx \f_\yy>$ or $<n_\xx n_\yy>$ as
$|\xx-\yy|\to \io$.  The situation could change at large coupling, where
the ground state could again be ordered: but this is outside the domain
of applicability of our techniques. The maximum value of $\l$, $|\l|<
\l_m(\s_0)$, for which we can prove \equ(1.7) goes to zero as
$\s_0\to\io$.

The problem of fermions with spin is, of course, much more interesting.
By the same argument given below this case can also be reduced to the
problem of fermions with spin, interacting with a short range potential.
However the fermionic interaction is attractive: and while this has no
consequences in the spinless case it has profound consequences in the
case with spin.  Unfortunately the spinning case is not understood in
the sense needed here to draw any conclusion, see [BM].
\pagina
\vskip1.cm
\0{\it\S 2 Reduction to a continuum problem.}
\vskip0.5cm
\numsec=2\numfor=1\pgn=1
When $\l=\n=0$ the phonon and fermion fields are independent. Their
respective Schwinger functions (imaginary time Green functions, see [FW]
for definitions) can be computed via the Wick rule from the two point
functions $S_\L(x)$ and $g_\L(x)$, where $\L\= [0,\b]\times
(-\fra{L}2,\fra{L}2]$ and $x=(t,\xx)\in\L$:
$$\eqalign{
S_\L(x)=&\fra1{\b L}\sum_{e^{i k_0\b}=1}\sum_{e^{i\kk L}=1,|\kk|<\p}
\fra{e^{-i k_0t-i\kk\xx}}{\s_0^2 k_0^2+1+ b^2 e_B(\kk)}\cr
g_\L(x)=&\fra1{\b L}\sum_{e^{i k_0\b}=-1}\sum_{e^{i\kk L}=1,|\kk|<\p}
\fra{e^{-i k_0t-i\kk\xx}}{-ik_0+e_F(\kk)}\cr}\Eq(2.1)$$
where $k=(k_0,\kk)$, and:
$$e_B(\kk)=2(1-\cos\kk),\qquad e_F(\kk)=(\cos \pf-\cos \kk) \Eq(2.2)$$
The summation rule over $k_0$ is to take the limit of $\sum_{|k_0|<N}$
as $N\to\io$, and we suppose that $\pf=\fra{2\p}{L}(n_F+\fra12)$, with
$n_F$ integer, so that $e_F(\kk)\ne0$ for all $\kk$.

An application of Trotter's formula allows us, as usual (see [BG1]), to
write an expression for the ground state infinite volume Schwinger
functions of the interacting fermions, $G(x)$, in terms of the gaussian
integral $P_B(d\F)$ with propagator $S_\L$ in \equ(2.1) and of the
grassmannian integral $P_F(d\psi)$ with propagator $g_\L$ in \equ(2.1);
for example:
$$\eqalign{
G(x)=&\lim_{\b\to\io\atop L\to\io}
\fra{\ig P_B(d\F)P_F(d\psi) e^{-V_\L(\F,\psi)}\psi^-_x\psi^+_0}
{\ig P_B(d\F)P_F(d\psi)e^{-V_\L(\F,\psi)}}\cr
V_\L(\F,\psi)=&\ig_0^\b
dt\sum_{\xx\in (-L/2,L/2]}(-\bar\n+\l\F_x)\psi^+_x\psi^-_x\cr
\bar\n =& \fra12(\n-\fra{\l^2}4) \hat S_\L(0) \cr}\Eq(2.3)$$
where the fields $\F$ satisfy periodic boundary conditions in $\xx$ and
$t$, while the $\psi$ fields (which are grassmannian variables) satisfy
periodic boundary conditions in $\xx$ and are antiperiodic in
$t$.\footnote{${}^1$}{\nota One should not confuse the fields $\F$
(which are real valued fields) with the operators $\f_x$; nor should one
confuse the grassmannian variables $\psi$ with the operators $a_x$.}
The term $(\l^2/2)\hat S_\L(0) \psi^+_x\psi^-_x$ in the definition of
$V_\L$ is produced by the shift of the field $\F_x$ which makes it
possible to write the measure $P_B(d\F)$ as a measure with zero mean.

The propagator $S(x)$ obtained from $S_\L(x)$ in the limits $\b\to\io,\,
L\to\io$ is exponentially decreasing; in fact, if $|t|<\b/2$ and
$|\xx|<L/2\;$, one can easily see that\footnote{${}^2$}{\nota Recall
that the boundary conditions are periodic.}:
$$|S_\L(x)| \le \fra{C(b)}{\s_0} e^{-\k_1|t|\s_0^{-1}} e^{-\k_2(b) |\xx|}
\Eq(2.4)$$
where $\k_1<1$ is independent of $b$, and:
$$ \k_2(b) = \cases { O(\log
b^{-1}) & for $b\to 0$ \cr O(b^{-1}) & for $b\to\infty$ \cr} ,\qquad
C(b) = \cases{ O(1) & for $b\to 0$ \cr O(b^{-1} \log b) & for
$b\to\infty$ \cr} \Eq(2.4a)$$
In particular, for $b=0$ it is, in the limit $\b\to\io$,
$S(x)= (2\s_0)^{-1} e^{-|t|\s_0^{-1}}\d_{\xx 0}$.

The integral over the bosonic field $\F$ is a gaussian integral, which
can be performed explicitly, yielding:
$$G(x)=\lim_{\b\to\io\atop L\to\io}
\fra{\ig P_F(d\psi) e^{-V_\L(\psi)}\psi^-_x\psi^+_0}
{\ig P_F(d\psi) e^{-V_\L(\psi)}}\Eq(2.5)$$
and, denoting $\ig_\L dx\cdot\=\sum_{\xx\in (-L/2,L/2]}\ig_0^\b dt \cdot$:
$$V_\L(\psi)= -\bar\n\ig_\L dx \psi_x^+\psi_x^--\fra{\l^2}8\ig_\L dx dy
\,S(x-y)\psi^+_x\psi^-_x\psi^+_y\psi^-_y\Eq(2.6)$$

We see that the problem has the same formal structure as that of
a fermion system on the continuous line interacting via a
short range potential, considered in [BGPS], see [G1] for a summary.
The attractivity is due to the positive definitness of $S(x)$, see
\equ(2.1).

Let us point out the main differences:
\*
\0(1) In formula \equ(2.3) the propagator $g(x)$ for the grassmannian
integral, defined in \equ(2.1) and \equ(2.2), has a different
``dispersion relation''; in[BGPS] the dispersion relation is
$e_F(\kk)=\fra12(\kk^2-\pf^2)$, while in the present case
$e_F(\kk)=(\cos \pf-\cos \kk)$. The difference is due to the fact that
in the Holstein model the fermions are on a lattice.

\0(2) The fermion potential \equ(2.6) is non local in time, while in
space it may even have zero range (in the case $b=0$, see \equ(2.4a)).
In the continuum problem considered in [BGPS] the potential had the
form \equ(2.6) with $v(\xx-\yy)\d(t-t')$ replacing $S(x-y)$, with
$x=(t,\xx), y=(t',\yy)$.
\*
These changes are of no consequence: in fact what was really used in
[BGPS] was that $\fra{d e_F}{d\kk}(\pf)>0$ which is true in the present
case as well, provided $0< \pf <\p$, as we suppose (``positive density
smaller than close packing'').

Therefore we can perform the decomposition of the propagator in the same
way as in [BGPS], see (13) in [G1], by writing:
$${1\over -i k_0+e_F(\kk)} = \fra{1-e^{-k_0^2+e_F(\kk)^2}}{-i k_0+e_F(\kk)}
+\fra{e^{-k_0^2-e_F(\kk)^2}}{-i k_0-e_F(\kk)}\Eq(2.7)$$
\equ(2.7) and \equ(2.1) generate a decomposition of the propagator
into a sum of an {\it ultraviolet part}, $g_{u.v.}(x)\=$ $g^{(>0)}(x)$,
and of an {\it infrared part}, $g_{i.r.}(x)\=$ $g^{(\le0)}(x)$.  The
decomposition \equ(2.7) allows us to represent $\psi^\pm_x$ as sums of
two independent grassmannian variables, that we denote $\psi^{(>0)}_x$
and $\psi^{(\le0)}_x$ with respective propagators $g^{(>0)}$ and
$g^{(\le0)}$.

The integration over $\psi^{(>0)}$ can be controlled by perturbation
expansions, as in [BGPS], and we shall integrate over it. The remaining
integral for the evaluation of the partition function, \ie for the
denominator of \equ(2.5), then becomes:
$$\ig P_F(d\psi^{(\le0)}) e^{- V^{(0)}(\psi^{(\le0)})} \Eq(2.8)$$
with:
$$\eqalign{
V^{(0)}(\psi)=&
-\fra{\l^2}2\ig dx\ dy\, S(x-y)\,\psi^+_x\psi^+_y\psi^-_y\psi^-_x+\cr
&-\l^2\ig dx dy\, S(x-y) g^{(>0)}(y-x)\psi^+_x\psi^-_y+\cr
&-(\bar\n +\l^2 g^{(>0)}(0))\ig dx\,\psi^+_x\psi^-_x+
\sum_{n=1}^\io\ig dx_1\ldots dx_{2n}\cdot\cr
&\cdot W_{2n}(x_1\ldots
x_{2n})\psi^+_{x_1}\ldots\psi_{x_n}^+\psi_{x_{n+1}}^-\ldots\psi^-_{x_{2n
}}\cr}\Eq(2.9)$$
where the kernels $W_{2n}(x_1\ldots)$ are analytic in $(\l^2,\bar\n)=r$
for $|r|<\e$ (for some $\e>0$) and verify the short range property:
$$\ig dx_1\ldots dx_{2n}\,|W_{2n}(x_1\ldots x_{2n})|\,
e^{\k d(x_1,\ldots,x_{2n})}< |\L| (D |r|)^{\max(2,n-1)}\Eq(2.10)$$
where $\k=\fra12\min(\k_1 \s_0^{-1}, \k_2(b))$, see \equ(2.4), and
$d(x_1,\ldots)$ is the length of the shortest path connecting all the
points $x_1,\ldots$ (regarded as points on the torus $\L$).

The main result achieved by \equ(2.9) is that we have ``disposed'' of
the ultaviolet part of the problem (in the evaluation of the partition
function) and we can say that the problem is reduced to an essentially
identical one with a purely infrared propagator and a new $V^{(0)}$
which, to lowest order in the couplings $r=(\l^2,\bar\n)$ has the same form
as the original one, as far as the quartic part is concerned, and a
slightly different (non local) quadratic part; plus ``higher order terms''
of every degree in the $\psi$ fields.  All the terms in \equ(2.9) are
well defined, and have a convergent power series in $r=(\l^2,\bar\n)$ if $r$
is small enough.

The proof of the above statements is simpler than the corresponding one
in [BGPS], see theorem 1, because in the present case there is a natural
ultraviolet cut off, at least in the $\xx$ direction, so that no
multi-scale decomposition of $g_{u.v.}(x)$ is needed.

The non locality of the quartic part in $V^{(0)}$ is not important even
for the infrared problem as it does not affect the notion of relevant or
marginal and irrelevant operators, which is the notion on which the
infrared integration is based when performed via the renormalization
group method in [BGPS]. Hence we reach the same conclusions about the
partition function and about the Schwinger functions (whose analysis can
be performed once the partition function can be estimated in detail, see
[BGPS], \S5 and \S6). The beta function is essentially $0$ (as shown in
\S7 of [BGPS]) and this allows us to draw the "same" conclusions as in
[BGPS].

In particular one can prove the anomalous asymptotic behaviour \equ(1.7) of
the two point Schwinger function with a coefficient $\h(\l)$ which is
analytic in $\l^2$ and in general is of order $\l^4$. However, if $b=0$,
one can see by an explicit calculation that the leading term in the
expansion of $\h(\l)$ vanishes, while the following one seems different
from zero.

Another interesting observation is that the formal ``Luttinger Theorem''
[LW] is valid in our case: it states that the density $\r$ is a function
only of $\pf$, that is $\r=\pf/\p$ for any $\l$: see appendix for a
discussion of this formal result (for which it would be nice to have a
rigorus proof).

In the next section we give some details on the recursion procedure
that we follow to obtain the estimates, and on how we define the
relevant operators: this will follow closely [BGPS] but should be
useful to the readers who wish to see where one is going, before
plunging themselves into the analysis of our estimates.  But neither the
estimates nor the proof of the vanishing of the beta function will be
reproduced here: they are identical to the corresponding proofs in
[BGPS].
\vskip1.cm
\0{\it\S3 The recursive evaluation of the partition function.}
\vskip0.5cm
\numsec=3\numfor=1\pgn=1
The following is the "standard" anomalous dimension renormalization
procedure. It is illustrated in quite a general context, in which one is
given \ap an arbitrary sequence of "wave function renormalization"
constants: $Z_0\=1,Z_{-1},Z_{-2},\ldots$.

The infrared propagator $g^{(\le 0)}(x)$ is decomposed as:
$$g^{(\le0)}(x)=\ig
e^{-ikx}\fra{dk}{(2\p)^2}\fra{T_0(k)}{-ik_0+e_F(\kk)}+
\ig
e^{-ikx}\fra{dk}{(2\p)^2}\fra{t_{-1}(k)}{-ik_0+e_F(\kk)}\Eq(3.1)$$
where $t_h(k)= \exp(-2^{-2h}(k_0^2+ e_F(\kk)^2))$ and
$T_0(k)=t_0(k)-t_{-1}(k)$.

Eq. \equ(3.1) allows us to represent $\psi^{(\le0)}$ as
$\psi^{(0)}+\psi^{(\le -1)}$ where $\psi^{(\le-1)}$ has propagator
$Z_0^{-1}t_{-1}(k) (-ik_0+e_F(\kk))^{-1}$ and $\psi ^{(0)}$ has a
propagator given by $Z_0^{-1}T_0(k)(-ik_0+e_F(\kk))^{-1}$. Let $\lis
P_{Z_0}(d\psi^{(0)})$ and $P_{Z_0}(d\psi^{(\le-1)})$ denote the
corresponding integrations, the grassmannian integral in \equ(2.8) thus
becomes:
$$\lis P_{Z_0}(d\psi^{(0)}) \, P_{Z_0}(d\psi^{(\le-1)})\Eq(3.2)$$
Using the sequence $Z_j$ one can then write the identity:
$$(Z_0 t_{-1}^{-1}+ Z_{-1}-Z_0)^{-1}= Z_{-1}^{-1} t_{-2}+
Z_{-1}^{-1}\lis\G^{(-1)}\Eq(3.3)$$
defining $\lis\G^{(-1)}(k)$.

Hence if $\psi^{(-1)}$ is a grassmannian field
with propagator:
$$\fra1{Z_{-1}}\fra{\lis \G^{(-1)}(k)}{-ik_0+e_F(\kk)}\Eq(3.4)$$
and $\psi^{(\le-2)}$ is a grassmannian field with propagator
$Z_{-1}t_{-2}(k)(-ik_0+e_F(\kk))^{-1}$, independent of $\psi^{(-1)}$ and
of $\psi^{(0)}$, then the grassmannian integral \equ(3.2) becomes
(formally), up to normalization constants:
$$\eqalign{
&\lis P_{Z_0}(d\psi^{(0)}\lis P_{Z_{-1}}(d\psi^{(-1)})
P_{Z_{-1}}(d\psi^{(\le-2)})\cdot\cr
&\cdot e^{(Z_{-1}-Z_0)\ig \psi^{(\le-1)+}_x
(\dpr_t+ e(i\dpr_\xx))\psi^{(\le-1)-}_x}\cr}\Eq(3.5)$$
as is easily checked with some algebra, if $\psi^{(\le-1)}=\psi^{(-1)}+
\psi^{(\le-2)}$.

By iteration:
$$\eqalign{
&P(d\psi^{(\le0)})=\left(\prod_{j=0}^h\lis
P_{Z_j}(d\psi^{(j)})\right)\cdot
P_{Z_h}(d\psi^{(\le h-1)})\cdot\cr
&\cdot\exp\sum_{j=0}^{h+1} (Z_{j-1}-Z_j)\ig \psi^{(\le j-1)+}_x
(\dpr_t+ e(i\dpr_\xx))\psi^{(\le j-1)-}_x\,dx\cr}\Eq(3.6)$$
and $\psi^{(\le0)}=\psi^{(0)}+\psi^{(-1)}+\ldots+\psi^{(h)}+\psi^{(\le
h-1)}$, with $\psi^{(j)}$ having propagator $Z_{j}^{-1}\lis \G^{(j)}(k)$
$(-ik_0+e_F(\kk))^{-1}\=$ $Z_{j}^{-1}\lis g^{(j)}$ with:
$$\lis \G^{(j)}(k)=t_j-t_{j-1}+ (1-t_j) t_j\,\fra{z_j}{1+z_j
t_j}\Eq(3.7)$$
if $z_j$ is defined by: $Z_j\=(1+z_j) Z_{j+1}$.

The integration \equ(2.8) can therefore be performed recursively by
setting:
$$\eqalign{
&e^{-V^{(h)}(\sqrt{Z_h}\psi^{(\le h)})}=
\ig \prod_{h'=h+1}^0\lis P_{Z_{h'}}(d\psi^{(h')})\cr
&e^{-V^{(0)}(\sqrt{Z_0}\psi^{(\le0)}+\sum_{h'=h}^{-1}(Z_{h'}-Z_{h'+
1})\ig dx\,\psi_x^{(\le h')+}(\dpr_t+ e(i\dpr_\xx))\psi^{(\le
h')-}_x}\cr}\Eq(3.8)$$
so that:
$$\eqalign{
&\ig e^{-V^{(0)}(\psi^{(\le0)})}P(d\psi^{(\le0)})=
\ig e^{-V^{(0)}(\sqrt{Z_0}\psi^{(\le0)})}\,P_{Z_0}(d\psi^{(\le0)})=\cr
&=\ig \lis P_{Z_h}(d\psi^{(h)})P_{Z_h}(d\psi^{(<h)})
e^{-V^{(h)}(\sqrt{Z_h}\psi^{(\le h)})}\cr}\Eq(3.9)$$
where $\psi^{(\le h)}=\psi^{(h)}+\psi^{(\le h-1)}$.

The (independent) "fluctuation fields" $\psi^{(h)}$ have propagators
with good scaling properties if $|z_h|<\fra12$: they can be represented
via quasi particle fields (see below) with propagators bounded,
unformly in $h$, by $2^h \g(2^hx)$ for some function $\g(x)$ which decays
exponentially fast as $x\to\io$.

The idea is to select the sequence $Z_h$ so that $|z_h|<\fra12$ and so
that the potential $V^{(h)}$ does not contain certain terms which would
otherwise be difficult to control.

To understand the choice of $Z_h$ one has to define the {\it relevant}
and {\it marginal} terms in $V^{(h)}$. Such notions are not naturally
defined from the $V^{(h)}$ considered as functions of the {\it particle
fields} $\psi^{(\le h)}$. They are very natural if $V^{(h)}$ is regarded as a
function of certain auxiliary fields that we call {\it the quasi
particle fields}.

The infrared propagators $g^{(\le h)}(x)$ can be decomposed as:
$$\eqalign{
g^{(\le h)}(x)=&\sum_{\oo=\pm1} e^{-i \pf\oo\xx} g^{(\le
h)}(x,\oo)\cr
g^{(\le h)}(x,\oo)=&e^{i
\pf\oo\xx}\ig_{-\io}^\io \fra{dk_0}{2\p} \ig_{-\p}^\p\fra{dk}{2\p}
\fra{t_h(k)\chi(2^{-h}\oo\kk)}{-ik_0+e_F(\kk)} e^{-ikx}\cr}\Eq(3.10)$$
where $\chi(r)$ may be chosen as $\p^{-1/2}\ig_{-\io}^r e^{-s^2}ds$, so
that $\chi$ is a smooth version of the step function, with the property
that $\chi(r)+\chi(-r)\=1$.

It is easy to check that, in the infinite (space-time) volume limit,
$g^{(\le h)}(x,\oo)$ is essentially scale
independent, \ie $h$ independent, as $h\to-\io$. In fact:
$$\eqalign{
&g^{(\le h)}(x,\oo)\simeq_{h\to-\io} 2^h \hat g(2^h x,\oo)\cr
&\hat g(x,\oo)=\ig_{-\io}^\io \fra{dk_0}{2\p} \ig_{-\io}^\io \fra{d\kk}{2\p}
\,\fra{t(k_0^2+(\sin \pf)^2\kk)^2)}{-ik_0+\oo\kk\sin \pf} e^{-ikx}\cr}
\Eq(3.11)$$
The decomposition \equ(3.10) of the covariance allows to perform a
related decomposition of the fields $\psi^{\pm (\le h)}_x$:
$$\psi^{\pm (\le h)}_x = \sum_{\oo=\pm1} e^{\pm i \pf\oo\xx} \psi^{\pm
(\le h)}_{x\oo} \Eq(3.11a)$$
The fields $\psi^{\pm (\le h)}_{x\oo}$, which will  be called the
{\it quasi particle fields}, are independent fields with propagator
$g^{(\le h)}(x,\oo)$; moreover they are antiperiodic also in the $\xx$
variable, so that we can write:
$$\psi^\pm_{x\oo}=\sum_{e^{i k_0\b}=-1,e^{i\kk L}=-1} e^{\pm ikx}
\psi^\pm_{k\oo} \Eq(3.17)$$

One can think that the $\psi^{(\le h)}_{x,\oo}$ have a distribution
which, ``up to scaling'', is $h$ independent. This means that the
distribution of $\psi^{(\le h)}$ is the same as that of
$2^{h/2}\hat\psi_{2^h x,\oo}$ where $\hat\psi_{x,\oo}$ has a propagator
$\hat g(x,\oo) \d_{\oo\oo'}$, see \equ(3.11) (up to corrections
vanishing fast as $h\to-\io$).
The fluctuation fields can also be expressed in terms of quasi particle
fields in the same way: their propagators have the good scaling
properties mentioned above.

Thus if we think of the infrared problem as that of integrating over the
quasi particle fields $\psi^{(\le h)}_{x\oo}$, we have a natural way of
introducing the notions of relevant, marginal and irrelevant operators.
Precisely we define the relevant operator to be:
$$F_2=\sum_{\oo,\oo'}\ig \psi^+_{x\oo}\psi^-_{x\oo'}
e^{i(\oo-\oo')\pf\xx} dx=\ig \psi^+_x\psi^-_x dx\Eq(3.12)$$
and the marginal operators as:
$$\eqalignno{
F_4=&\ig \psi^+_{x+}\psi^+_{x-}\psi^-_{x-}\psi^-_{x+}\,dx&\eq(3.13)\cr
F_{2,\t}=&\sum_{\oo,\oo'}\ig
e^{i(\oo-\oo')\pf\xx}\psi^+_{x\oo}(\dpr_t +i\oo' \sin \pf
\DD_{\oo'})\psi^-_{x\oo'}\,dx=
\ig\psi^+_{x}(\dpr_t+e_F(i\dpr_\xx))\psi^-_x\,dx\cr
F_{2,\s}=& \sum_{\oo,\oo'}\ig
e^{i(\oo-\oo')\pf\xx}\psi^+_{x\oo}\ i\oo'\sin \pf
\DD_{\oo'} \psi^-_{x\oo'}\,dx=
\ig\psi^+_{x}e_F(i\dpr_\xx)\psi^-_x\,dx\cr}$$
where $\DD_\oo$ is a suitable operator, which acts by
multiplication by $M_\oo(\kk)$ on the Fourier transforms, with
$$M_\oo(\kk)= -2i\sin \kk/2 \fra{\sin(\pf+\oo\kk/2)}{\sin \pf}
\simeq_{\kk\to 0} -i\kk \Eq(3.14)$$
Note that, while the second degree operators can be expressed easily in
terms of the particle fields, the same is not true for the $F_4$.

The ``usual'' power counting attributes a size to the above operators
evaluated by extending the integral over a box of size $2^{-h}$, \ie of
volume $2^{-2h}$, and by attributing to each field a size $2^{h/2}$ as
suggested by the scaling properties discussed above; furthermore each
derivative contributes to the size a factor $2^h$.

Hence the conventional power counting attributes to $F_2$ a size that is
evaluated as $2^{-2h}(2^{h/2})^2=2^{-h}$ (hence $F_2$ is relevant). The
size of $F_4$ is $2^{-2h}(2^{h/2})^4=1$ and $F_{2,\t},F_{2,\s}$ have
size $2^{-2h}2^h(2^{h/2})^2=1$ (hence they are marginal). All the other
local operators are irrelevant (\ie they have sizes $2^{h/2}$ or less).

Note that the size of the various operators is clear if they
are regarded as functions of the quasi particle fields.

Given $V^{(h)}(\psi)$, we must identify the relevant and marginal parts
of $V^{(h)}$. This is done by introducing the {\it localization
operator} $\LL$: it is a linear projection operator which is $0$ unless
acting on a fourth degree or on a second degree monomial in the fields.
Imagining that $V^{(h)}$ is expressed in terms of quasi particle fields,
hence as a sum (or integral) of monomials in the quasi particle fields,
the action of $\LL$ on the fourth degree monomials is described by:
$$\LL \,\psi^+_{x_1\oo_1}\psi^+_{x_2\oo_2}\psi^-_{x_3\oo_3}\psi^-_{x_4\oo_4}
=\fra12\sum_{i=1}^2
\psi^+_{x_i\oo_1}\psi^+_{x_i\oo_2}\psi^-_{x_i\oo_3}\psi^-_{x_i\oo_4}
\Eq(3.15)$$
while the action of $\LL$ on the second degree monomials is:
$$\LL \psi^+_{x\oo_1}\psi^-_{y\oo_2}=
\psi^+_{x\oo_1}[ \psi^-_{x\oo_2}+h_\b(y_0-x_0)\dpr_{x_0}\psi^-_{x\oo_2}
\psi^-_{x\oo_1}h_L(\yy-\xx)\DD_{\oo_2}\psi^-_{x\o_2}]\Eq(3.16)$$
where
$$h_M(s-t) \equiv {M\over 2i\p}[e^{i\p(s-t)/M}-e^{-i\p(s-t)/M}]
\Eq(3.16a)$$
is an antiperiodic approximation of $(s-t)$, which converges to it as
$M\to\io$.

The operator $\DD_\oo$ differs from the analogous operator defined in
[BGPS], where $\xx$ was a continuum variable. However, it is easy to see
that one has to change the bounds obtained in [BGPS] only in minor
points, without affecting the final results. In \equ(3.16) we have also
taken explicitely into account the boundary conditions, while in [BGPS]
this problem was neglected. However, as explicitely shown in the
analysis of the spinning case in [BM], this approximation does not play
any relevant role.

\vskip1cm\numsec=1\numfor=1\pgn=1
\0{\it Appendix}
\vskip.5cm

In this appendix we give a heuristic derivation of the so called
Luttinger Theorem, following the analysis presented in [LW]. We believe
that the techniques used in [BGPS] to study the Schwinger functions
could be used also to prove rigorously the Luttinger Theorem, but this
has not been done yet.

We consider a system in finite volume $L$ at zero temperature. If we
define the hamiltonian $H$ so that the ground state $\ket{0}$ has
eigenvalue $0$, the two point Schwinger function is defined in the
following way:

$$G(\xx,t) = \c(t>0) \bra{0} a^-_\xx e^{-tH} a^+_0\ket{0}-
\c(t\le 0) \bra{0} a^+_0 e^{+tH} a^-_\xx\ket{0} \Eqa(a1)$$
where $\chi(A)$ is the characteristic function of the set of points
where the condition $A$ holds.

Hence its Fourier tranform $\hat G(\kk,k_0)$ is given by:
$$\eqalign{
\hat G(\kk,k_0) &= \int_0^\io \ dt \Big[ e^{ik_0 t} \bra{0} a^-_\kk e^{-tH}
a^+_\kk \ket{0} - e^{-ik_0 t} \bra{0} a^+_\kk e^{-tH}
a^-_\kk \ket{0}\Big]\cr
&= \sum_{n:E_n>0} \int_0^\io \ dt \Big[ e^{-t(E_n-ik_0)} |\bra{0} a^-_\kk
\ket{n}|^2 - e^{-t(E_n+ik_0)} |\bra{0} a^+_\kk\ket{n}|^2 \Big] =\cr
&= \sum_{n:E_n>0} \Big[ \fra{|\bra{0} a^-_\kk
\ket{n}|^2}{-ik_0+E_n} + \fra{|\bra{0} a^+_\kk\ket{n}|^2}{-ik_0-E_n} \Big]
\cr}\Eqa(a2)$$
where $E_n, n=0,\ldots$, are the eigenvalues of $H$ and $\ket{n}$ the
corresponding eigenstates. The ground state gives no contribution to the
sum in \equ(a2) (so allowing the integration over $t$) because of the
conservation of the total momentum, which implies that $\bra{0}a^\pm_\kk
\ket{0}=0$, when $\kk\not=0$, and because of a cancellation between the
two different terms in the r.h.s. of \equ(a2), when $\kk=0$.

The function $\hat G(k)$, $k=(\kk,k_0)$, is often written also in the
following form:
$$\hat G(k)= {1\over -ik_0 + e_F(\kk) +\Si(k)} \Eqa(a3)$$
which defines the {\it self-energy function} $\Si(k)$. As is well known,
[LW], $\Si(k)$ can be expressed in perturbation theory as the sum of all
connected graphs with two external lines, which are irreducible, that is
which do not become disconnected by erasing any internal line.

By using \equ(a1), one sees immediately that the density $\r$ is given by:
$$\r = -\lim_{t\to 0^-} G(0,t) = {1\over L}\sum_\kk
\lim_{\e\to 0^+} \int {dk_0\over 2\p} \fra{e^{ik_0\e}}{ik_0
-e_F(\kk)-\Si(k)}\Eqa(a4)$$

Let us write
$${1\over \z -e_F(\kk)-\Si(\kk,-i\z)} = {\dpr\over \dpr\z}\log [\z
-e_F(\kk)-\Si(\kk,-i\z)] + \fra{\dpr\Si(\kk,-i\z)/\dpr\z}{\z
-e_F(\kk)-\Si(\kk,-i\z)} \Eqa(a5)$$
by choosing the branch of $\log z$ so that $\log z = \log |z| +i
\arg{z}$, $0<\arg{z}<2\p$.

It was argued in [LW], if we insert the r.h.s of \equ(a5) in the r.h.s. of
\equ(a4), the second term gives no contribution, that is:
$${1\over L}\sum_\kk \lim_{\e\to 0^+} \int {dk_0\over 2\p} e^{ik_0\e}
\fra{\dpr\Si(k)/\dpr k_0}{ik_0 -e_F(\kk)-\Si(k)} = -{1\over L}\sum_\kk
\int {dk_0\over 2\p} \Si(k) {\dpr \hat G(k)\over \dpr k_0} = 0 \Eqa(a6)$$

We recall briefly the arguments given in [LW] to justify
\equ(a6). Indeed, our problem is not
explicitly considered in [LW], where the interaction between the
fermions is local in time; hence only the case $b=0$ is covered by [LW].
However, their arguments extend in a trivial way to our general case.

We start from the observation (not easy to prove rigorously) that
$\Si(k)$ can be thought as the sum of all graphs (connected and
irreducible) with two external lines, such that the internal lines carry
the complete propagator $\hat G(k)$ and the following condition is
satisfied: there is no proper subgraph with two external lines. We shall
call such graphs, as usual, {\it skeleton graphs} and say that a graph
is of order $n$, if it contains $n$ four-fermions interacting terms
(hence it has $2n$ vertices).

Let $\Si_n(k)$ be the sum of all skeleton graphs of order $n$ and let
$Y_n$ be the sum of all vacuum (that is with no external lines) skeleton
graphs of order $n$, which can be obtained by closing the two external
lines of a graph contributing to $\Si(k)$ with a propagator $\hat
G(k)$. Since each graph contributing to $Y_n$ can be obtained in $2n$
different ways from a graph contributing to $\Si(k)$, we have:
$$Y_n = {1\over 2n} {1\over L}\sum_\kk \int {dk_0\over 2\p} \Si(k)
\hat G(k)\Eqa(a7)$$
Moreover, given a graph ${\cal G}$ contributing to $Y_n$, its value
$Y_n^{\cal G}$ can be written in the following way:
$$\eqalign{
Y_n^{\cal G} &= \int dk^{(1)}\ldots dk^{(2n)} dq^{(1)}\ldots dq^{(n)}
\hat G(k^{(1)}) \cdots \hat G(k^{(2n)}) \cdot \cr
&\cdot S(q^{(1)})\ldots S(q^{(n)}) \prod_{i=1}^{2n} \d(
k^{(s_i)} -k^{(r_i)} + q^{(b_i)} )\cr} \Eqa(a8)$$
where $\int dk \= L^{-1}\sum_\kk
\int dk_0/(2\p)$ and $k^{(1)},\ldots, k^{(2n)}$ are the (space-time)
momenta of the $2n$ fermion lines, $q^{(1)},\ldots, q^{(n)}$ are the
momenta of the $n$ phonon lines and, finally, $k^{(s_i)}$ and
$k^{(r_i)}$ are the momenta of the fermion entering and leaving the
vertex $i$, respectively, while $q^{(b_i)}$ is the momentum of the
phonon propagator entering vertex $i$.  Moreover, if we consider the sum
of all the quantities that we obtain by substituting, in the r.h.s. of
\equ(a8), one fermion propagator by its derivative with respect to
$k_0$, we get:
$$\eqalign{
&\sum_{j=1}^{2n} \int dk^{(1)}\ldots dk^{(2n)} dq^{(1)}\ldots dq^{(n)}
\hat G(k^{(1)}) \cdots \fra{\dpr \hat G}{\dpr k_0^{(j)}}(k^{(j)}) \cdots
\hat G(k^{(2n)}) \cdot \cr
&\cdot S(q^{(1)})\ldots S(q^{(n)}) \prod_{i=1}^{2n} \d( k^{(s_i)}
-k^{(r_i)} + q^{(b_i)} ) =\cr &= \pm \sum_{j=1}^{2n} \int dk^{(1)}\ldots
dk^{(2n)} dq^{(1)}\ldots dq^{(n)}
\hat G(k^{(1)}) \cdots \hat G(k^{(2n)}) \cdot \cr
&\cdot S(q^{(1)})\ldots S(q^{(n)})
( \fra{\dpr}{\dpr k_0^{(s_j)}} +\fra{\dpr}{\dpr k_0^{(r_j)}} )
\prod_{i=1}^{2n} \d(k^{(s_i)} -k^{(r_i)} + q^{(b_i)} ) = 0\cr} \Eqa(a9)$$
If we now sum the l.h.s. of \equ(a9) over ${\cal G}$, we get, by an argument
similar to that used in order to prove \equ(a7), that:
$$2n \int dk\ \Si(k) \fra{\dpr \hat G}{\dpr k_0}(k) = 0 \Eqa(a10)$$
so that \equ(a6) is proved.

The formal identity \equ(a6) implies that $\r = {1\over L}\sum_\kk
\r_\kk$, with
$$\r_\kk =
\lim_{\e\to 0^+} \int_{-i\io}^{+i\io} {d\z\over 2\p i} e^{\e\z}
{\dpr\over \dpr\z} \log [\z -e_F(\kk)-\Si(\kk,-i\z)]\Eqa(a11)$$
Moreover, \equ(a2) implies that the integrand in the r.h.s. of \equ(a11)
is analytic in all the complex $\z$ plane, except on the real axis, where
there are branch points at $\z=\pm E_n$, $n>0$. Therefore, we can deform the
integration contour into the contour $\cal C$ of Fig. 1

\initfig{fig1abczwk}
\write13</cambio_coordinate{ 
\write13<4 copy exch pop exch sub  
\write13<6 1 roll exch pop sub    
\write13<4 1 roll translate       
\write13<2 copy exch atan rotate  
\write13<2 exp exch 2 exp add sqrt  
\write13<} def>
\write13</freccia { gsave 
\write13<cambio_coordinate 
\write13<dup 0 0 moveto 0 lineto 
\write13<2 div 0 translate>
\write13<15 rotate 0 0 moveto -5 0 lineto -30 rotate 0 0 moveto -5 0 lineto>
\write13<stroke grestore } def>
\write13<>
\write13</frecciafin { gsave 
\write13<cambio_coordinate 
\write13<dup 0 0 moveto 0 lineto 
\write13<0 translate>
\write13<15 rotate 0 0 moveto -5 0 lineto -30 rotate 0 0 moveto -5 0 lineto>
\write13<stroke grestore } def>
\write13<>
\write13</sfreccia { gsave 
\write13<setlinewidth freccia grestore} def>
\write13<>
\write13<20 50 280 50 frecciafin 150 10 150 90 frecciafin>
\write13<20 30 150 30 2 sfreccia 150 30 150 50 2 sfreccia>
\write13<150 50 150 70 2 sfreccia 150 70 20 70 2 sfreccia>
\write13<stroke>
\endfig

\vfill\eject
\insertplot{300pt}{100pt}{%
\ins{155pt}{35pt}{$-\d$}\ins{155pt}{75pt}{$+\d$}%
\ins{275pt}{45pt}{${\rm Re\,}\z$}\ins{155pt}{95pt}{${\rm Im}\,\z$}%
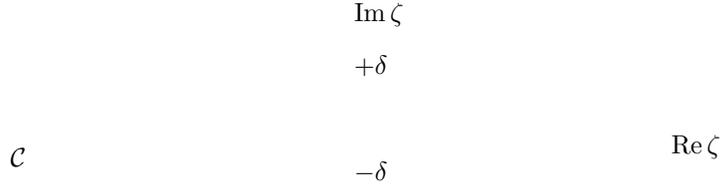
\ins{25pt}{40pt}{${\cal C}$}}{fig1abczwk}{Fig. 1: Integration
Contour for \equ(a11).}

\vskip.3truecm
\noindent and then we can integrate by parts. We get:
$$\eqalign{
\r_\kk &= \lim_{\h\to 0^+} \fra{\log [-e_F(\kk)-\Si(\kk,-\h)] -
\log[-e_F(\kk)-\Si(\kk,\h)]}{2\p i} \ \cr
&-\ \lim_{\e\to 0^+} \e\int_{\cal C} {d\z\over 2\p i} e^{\e\z}
\log [\z -e_F(\kk)-\Si(\kk,-i\z)] \cr} \Eqa(a12)$$
It is easy to see that the second term in \equ(a12) vanishes, if the
argument of the logarithm has a negative real part for ${\,\rm Re\,}\z
\to-\io$, by recalling that we have chosen the branch of $\log z$ with
the cut along the positive real axis.

We want to show that this is certainly true, if $|\bra{0} a^+_\kk
\ket{n}|\to 0$ sufficiently fast, as $E_n\to\io$. We note that ${\,\rm
Re\,} [\z -e_F(\kk)-\Si(\kk,-i\z)]$ has the same sign as $-{\,\rm Re\,}
\hat G(\kk,-i\z)$; moreover, by \equ(a2), if $\z=u+i\d$:
$$-{\,\rm Re\,} \hat G(\kk,-i\z) = \int_{-\io}^{+\io} \s_\kk(dE)
\fra{u-E}{(u-E)^2+ \d^2}\Eqa(a12a)$$
where $\s_\kk(dE)$ is a probability measure, as it is easy to check.
Moreover, if $u\le -u_0$, with $u_0$ large enough, there is an interval
$[E_1,E_2]$, such that $|u-E|\ge |u|/2$ for $E\in [E_1,E_2]$ and
$\int_{[E1,E_2]} \s_\kk(dE) =1/2$; hence it is easy to show that:
$$-{\,\rm Re\,} \hat G(\kk,-i\z) \le (2\d)^{-1} \int_{-\io}^{u} \s_\kk(dE)
- \fra{|u|}{u^2+ 4\d^2}\Eqa(a12b)$$
which immediately implies that $-{\,\rm Re\,} \hat G(\kk,-i\z)$ is definitely
negative for $u\to -\io$, if
$$\lim_{u\to-\io} u \int_{-\io}^u \s_\kk(dE) = 0 \Eqa(a12c)$$

Since, by \equ(a2), the function $\Si(\kk,-i\z)$ is real at $\z=0$,
it follows from \equ(a12) that:
$$\r = {1\over L}\sum_\kk \c[e_F(\kk)+\Si(0,\kk) <0] \Eqa(a13)$$

Equation \equ(a13) implies that, in the limit of infinite
volume, there is a relation, independent of the strength of the
interaction $\l$, between the density and the Fermi momentum $\pf$,
defined as the value of $\kk$, such that $e_F(\kk)+\Si(\kk,0)=0$ (that is
the value of $\kk$ where the interacting propagator is singular for
$k_0=0$). For $\l=0$ we have $\r= \pf/\p$, hence this relation has to be
valid for any $\l$.
\pagina
{\it Acknowledgements:} We thank Nicolas Macris for useful comments.
This work has been supported supported in part by NSF Grant DMR
92--13424 4--20946, by CNR-GNF and by MURST 40\%.  It is also part of
the research program of the European Network on: "Stability and
Universality in Classical Mechanics", \# ERBCHRXCT940460.

\vskip2.cm
\0{\it References}
\vskip0.5cm

\0[BG1] Benfatto, G., Gallavotti, G.: {\it Perturbation theory of the
Fermi surface in a quantum liquid. A general quasi particle formalism
and one dimensional systems}, Journal of Statistical Physics, {\bf 59},
541--664, 1990.

\0[BGPS] Benfatto, G., Gallavotti, G., Procacci, A., Scoppola, B.:
{\it Beta function and Schwin\-ger functions for a many fermion system in
one dimension. Anomaly of the Fermi surface.}, Communications in
Mathematical Physics, {\bf 160}, 93--172, 1994.

\0[BM] Bonetto, F., Mastropietro, V.: {\it Renormalization group theory
in a $d=1$ system of interacting fermions in a periodic potential},
Communications in Mathematical Physics, 1995.

\0[FW] Fetter, A.L. and Walecka, J.D.:  {\it Quantum Theory of Many
Particle Systems}, McGraw Hill, N.Y., 1971.

\0[G1] Gallavotti, G.: {\it One dimensional anomaly of the Fermi surface},
in "On Three levels", ed. M. Fannes, C. Maes, A. Verbeure, NATO ASI
series B, vol. 324, Plenum Press, New York, 1994, p. 165--173.

\0[L] Luttinger, J.M.: {\it Fermi surface and some simple equilibrium
properties of a system of interacting fermions}, Physical Review, {\bf
119}, 1153--1163, 1960.

\0[LW] Luttinger, J.M., Ward, J.C.: {\it Ground-state energy of a
many-fermion system.II}, Physical Review, {\bf 118}, 1417--1427, 1960.

\0[MM] Messager, A., Miracle--Sol\'e, S.: {\it Low temperature states in
the Falikov Kimball model}, preprint, Marseille, november 1994.

\ciao